\documentclass[twocolumn,astrosymb,times]{aastex7}

\usepackage{amsmath}
\usepackage{xspace}
\usepackage{natbib}
\usepackage{listings}
\usepackage{hyperref}

\renewcommand{\vec}[1]{\ensuremath{\mathbf{#1}}\xspace}
\newcommand{\ud}{\ensuremath{\mathrm{d}}}
\newcommand{\roadrunner}{\texttt{RoadRunner}\xspace}
\newcommand{\pytransit}{\texttt{PyTransit}\xspace}
\newcommand{\exoiris}{\texttt{ExoIris}\xspace}
\newcommand{\tsmodel}{\texttt{TSModel}\xspace}
\newcommand{\ldtkm}{\texttt{LDTkLDM}\xspace}
\newcommand{\ldtk}{\texttt{LDTk}\xspace}
\newcommand{\istar}{\ensuremath{I_\star}\xspace}
\newcommand{\aplanet}{\ensuremath{A_\mathrm{p}}\xspace}
\newcommand{\iplanet}{\ensuremath{\hat{I}_\mathrm{p}}\xspace}
\newcommand{\python}{\texttt{Python}\xspace}

\newcommand{\teff}{\ensuremath{T_\mathrm{eff}}\xspace}
\newcommand{\logg}{\ensuremath{\log g}\xspace}
\newcommand{\metal}{\ensuremath{[\mathrm{Fe}/\mathrm{H}]}\xspace}

\newcommand{\ldc}{\ensuremath{\mathbf{l}}\xspace}
\newcommand{\bias}{\ensuremath{O}\xspace}
\newcommand{\porb}{\ensuremath{\boldsymbol{\Theta}}\xspace}
\newcommand{\jwst}{JWST\xspace}
\renewcommand{\micron}{\ensuremath{\mu\text{m}}\xspace}

\newcommand{\logl}{\ensuremath{\log L}\xspace}
\newcommand{\covmat}{\ensuremath{\boldsymbol{\Sigma}}\xspace}
\newcommand{\meansp}{\ensuremath{\boldsymbol{\mu}}\xspace}

\shorttitle{ExoIris}

\begin{document}

\title{ExoIris: fast exoplanet transmission spectroscopy in Python}

\author[orcid=0000-0001-5519-1391, sname='Parviainen']{Hannu Parviainen}
\altaffiliation{Ram\'on y Cajal Fellow}
\affiliation{Departamento de Astrof\'isica, Universidad de La Laguna (ULL), E-38206 La Laguna, Tenerife, Spain}
\affiliation{Instituto de Astrof\'isica de Canarias (IAC), E-38205 La Laguna, Tenerife, Spain}
\email[show]{hannu@iac.es}

\begin{abstract}
I present \exoiris, a user-friendly Python package for exoplanet transmission and emission spectroscopy. Unlike existing tools, \exoiris models two-dimensional spectrophotometric transit time series directly and supports the joint analysis of multiple datasets obtained with different instruments and at different epochs, as well as modeling stellar spot crossings and the influence of unocculted heterogeneities (the transit light source effect). These features enable a self-consistent estimation of both wavelength-independent and wavelength-dependent parameters. They offer a more robust workflow than the commonly used two-step approach, in which a "white" light curve is fitted first, and the transmission spectrum is then derived from independent fits constrained by the white-light solution. Despite its increased flexibility and robustness, \exoiris remains computationally efficient. A low-resolution transmission spectrum can be estimated from a single \jwst NIRISS transit observation in $\sim5$~minutes assuming white noise, and in $\sim15$~minutes when accounting for time-correlated systematics using a Gaussian process noise model, on a standard desktop computer.
\end{abstract}


\section{Introduction} 
\label{sec:intro}

I present \exoiris,\!\footnote{The \exoiris package is available via PyPI and GitHub (\url{https://github.com/hpparvi/ExoIris}). Documentation and tutorials can be found at \url{https://exoiris.readthedocs.io}.} an open-source \python package for transmission and emission spectroscopy of exoplanets. \exoiris differs from existing tools by directly modeling two-dimensional spectrophotometric transit datasets (time series of spectroscopic transit or eclipse observations) and by supporting joint analyses of multiple datasets obtained with different instruments and at different epochs.

In \exoiris, wavelength-dependent quantities (planet-star radius ratio and stellar limb-darkening profile) and wavelength-independent quantities (e.g., orbital elements) are inferred jointly from the full dataset. This contrasts with the common two-step approach, in which each spectroscopic light curve is fitted independently, often using tight priors from white-light curve analysis and fixed limb-darkening coefficients from stellar atmosphere models. Such methods fail to fully exploit the information content of the observations and may underestimate uncertainties. By modeling the full two-dimensional dataset, \exoiris enables self-consistent inference of all relevant parameters, yielding more robust parameter estimates and transmission spectra.

\exoiris also supports joint modeling of heterogeneous datasets, allowing users to account for instrument- and time-dependent effects such as transit timing variations (TTVs) and instrumental systematics \citep[e.g.][]{Moran2023, Alderson2025}. Time-correlated noise and systematics can be modeled with Gaussian processes (GPs), using fixed or inferred hyperparameters. The model can also include an arbitrary number of stellar spot crossings, as well as time-varying levels of contamination from unocculted spots and faculae \citep[the transit light source effect, TLSE;][]{Rackham2018, Rackham2019, Rackham2024}.

Despite its flexibility, \exoiris remains computationally efficient. The planet-star radius ratio is modeled as a wavelength-dependent interpolating function defined by a set of freely placed knots, whose values are included as free parameters in the model, with several interpolation schemes available. This approach decouples the spectral resolution of the transmission spectrum from that of the data, enabling high-resolution sampling of sharp absorption features while allowing the smoother continuum to be modeled at lower resolution. The number of free parameters in the final model is primarily determined by the number of radius-ratio knots, and adding new datasets does not significantly increase the model's dimensionality if the wavelength space is already covered. For example, a joint fit to a single \jwst NIRISS transit with $\mathcal{R} \sim 300$, 100 radius-ratio knots, and a GP noise model typically completes in 10--20 minutes on a standard desktop computer. More complex analyses, such as the joint modeling of eight WASP-39~b Early Release Science datasets (Sect.~\ref{sec:example}), complete within a few hours.

\section{ExoIris}
\label{sec:exoiris}

\subsection{Spectrophotometry Model Overview}
\label{sec:exoiris.model}

An \exoiris transmission spectroscopy analysis considers a collection of spectrophotometric datasets, where each dataset contains spectrophotometric fluxes and uncertainties from a single transit observation, along with time and wavelength sampling and parameter grouping metadata. 
The observed spectrophotometric flux for dataset $d$, as a function of time $t$ and wavelength $\lambda$, is modeled as
\begin{equation}
    F_\mathrm{d}(t, \lambda) = \bias_\mathrm{d} + (1-\bias_\mathrm{d}) B_\mathrm{d}(t, \lambda)\; T\left(t, k(\lambda), \ldc(\lambda), \porb \right), \label{eq:full_model}
\end{equation}
where \bias is a flux offset term, $B$ is a baseline flux model, and $T$ is the spectrophotometric transit model. The model depends on the planet-star radius ratio $k$, the limb-darkening parameters \ldc, and the orbital parameter vector \porb (comprising the zero epoch $t_0$ for each epoch group, orbital period $p$, inclination $i$, eccentricity $e$, and argument of periastron $\omega$). 
Each dataset is assigned to offset, noise, epoch, and baseline model groups, which determine the model parameterization. These are described in detail in Sects.~\ref{sec:exoiris.offset}, \ref{sec:exoiris.noise}, \ref{sec:exoiris.ttvs}, and \ref{sec:exoiris.baseline}, respectively.

\begin{figure*}
    \centering
    \includegraphics[width=1\linewidth]{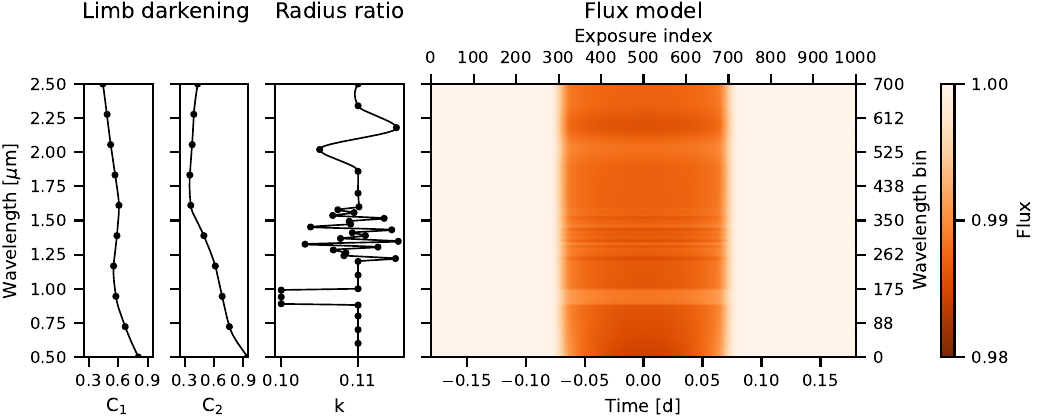}
    \caption{Example of a spectrophotometric transit model. The first two panels show the power-2 limb-darkening coefficients as splines over wavelength with freely placed knots. The third panel shows the planet's radius ratio modeled in the same way. The rightmost panel displays the model flux as a function of time and wavelength.}
    \label{fig:model_example}
\end{figure*}

The planet-star radius ratio, $k(\lambda)$, is modeled as a wavelength-dependent interpolating function, decoupling the resolution of the transmission spectrum from the spectroscopic resolution of the data (Sect.~\ref{sec:exoiris.radius_ratio}). The limb-darkening vector \ldc depends on the chosen model. Currently, \exoiris supports two-parameter analytical limb-darkening laws with coefficients modeled as splines and a numerical \ldtkm-based model (Sect.~\ref{sec:exoiris.limb_darkening}). In the analytical case, \ldc consists of limb-darkening coefficients at the knot wavelengths; in the numerical case, it contains the stellar effective temperature, metallicity, and surface gravity.

Figure~\ref{fig:model_example} illustrates a single spectrophotometric model for a dataset with 1000 exposures and 600 wavelength elements using cubic B-spline interpolation for radius ratio and limb-darkening coefficients. The model flexibly captures both sharp and smooth variations in $k$, allowing high resolution where needed and lower resolution elsewhere.

\subsection{Transmission Spectroscopy Transit Model}
\label{sec:exoiris.transit_model}

\exoiris uses the \tsmodel transit model from \pytransit v2.6 \citep{Parviainen2015, Parviainen2020a, Parviainen2020b}, specifically designed for transmission spectroscopy. It is based on the latest version of the \roadrunner model \citep{Parviainen2020b}, which enables fast light curve evaluation with arbitrary radially symmetric stellar surface brightness profiles. This flexibility supports both analytical limb-darkening laws and numerical profiles derived from stellar atmosphere models.

Assuming a radially symmetric stellar intensity profile $I(z)$, where $z$ is the projected distance from the stellar center normalized to the stellar radius, the in-transit flux is given by
\begin{align}
    I &= \int_{A_\star} I(z) \;\ud A - \int_{A_\mathrm{p}} I(z) \;\ud A \\
      &= 2\pi  \int_0^1 z\;I(z)\;\ud z -  \int_{A_\mathrm{p}} I(z) \;\ud A \\
      &= \hat{I}_\star A_\star - \hat{I}_\mathrm{p} A_\mathrm{p},
\end{align}
where $\hat{I}_\star$ is the mean surface brightness of the stellar disk, $A_\star$ its area, $\hat{I}_\mathrm{p}$ the mean surface brightness of the stellar disk occulted by the planet, and $A_\mathrm{p}$ the occulted area. If the stellar radius and out-of-transit flux are normalized to unity, the expression simplifies to
\begin{equation}
    I = \frac{\pi\hat{I}_\star - \hat{I}_\mathrm{p} A_\mathrm{p}}{\pi\hat{I}_\star} =  \frac{\istar - \hat{I}_\mathrm{p} A_\mathrm{p}}{I_\star},
\end{equation}
where $\istar = \pi \hat{I}_\star$ is the total stellar flux. 

Among these quantities, $\istar$ and $\aplanet$ are straightforward to calculate. The main computational challenge lies in evaluating $\hat{I}_\mathrm{p}$ efficiently for arbitrary limb-darkening profiles. The \roadrunner model addresses this by expressing $\iplanet$ as a weighted sum over the stellar intensity profile, reducing the model evaluation to a dot product between two one-dimensional arrays multiplied by the area occulted by the planet:
\begin{equation}
        I = \frac{\istar - \vec{w}(k, b) \cdot \vec{l} \times a(k, b)}{I_\star}, \label{eq:model}
\end{equation}
where $\vec{w}(k, b)$ is a weight vector dependent on the radius ratio $k$ and sky-projected planet-star center separation $b$ as described in \citet{Parviainen2020b}, $\vec{l}$ is the discretized limb-darkening profile, and $a(k, b)$ is the occulted area.

The mean occulted brightness $\vec{w} \cdot \vec{l}$ varies smoothly with $b$ for fixed $k$, and can be precomputed as a one-dimensional interpolation table to further speed up evaluation. The resulting model becomes
\begin{equation}
      I = \frac{I_\star - \hat{I}_\mathrm{p}(b) \times a(k, b)}{I_\star},
\end{equation}
where $\hat{I}_\mathrm{p}(b)$ is the precomputed mean occulted intensity for fixed $k$, interpolated as a function of $b$. The intersection area $a(k, b)$ is computed using the circle-circle intersection routine by \citet{Agol2020}, which also provides its derivatives with respect to $k$ and $b$.

Separating the limb-darkening term from the geometric occultation enables high computational efficiency, making the \roadrunner model well suited for transmission spectroscopy. A transmission spectroscopy dataset derived from a spectroscopic time series consists of flux estimates in multiple narrow passbands, all sampled at identical points in time. This allows for several optimizations that significantly reduce the computational cost of the model:
\begin{enumerate}
    \item The planet-star center distance, $b$, is independent of passband and only needs to be computed once per exposure.
    \item Although the weight vector $\vec{w}$ depends on $k$, the derivative $\partial \hat{I}_\mathrm{p} / \partial k$ is negligible compared to other effects of $k$ variation. Thus, the weight vector computed for the mean radius ratio $\hat{k}$ can be used across all passbands with minimal impact on accuracy.
    \item While computing the circle-circle intersection area is already inexpensive, it can be further accelerated. In the fully occulted regime ($b < 1 - \max(\vec{k})$, where $\vec{k}$ is the vector of radius ratios across passbands), the intersection area simplifies to
    \begin{equation}
    a(k, b) = \pi k^2.
    \end{equation}
    In the partial overlap regime ($1 - \max(\vec{k}) \leq b \leq 1 + \max(\vec{k})$), a first-order Taylor expansion around $\hat{k}$ gives
    \begin{equation}
        a(k, b) = a(\hat{k}, b) + (k-\hat{k}) \frac{\partial a}{\partial k}
    \end{equation}
    where $\partial a / \partial k = 2 \hat{k} \kappa_0(\hat{k})$, and $\kappa_0$ is defined in Eq.~(33) of \citet{Agol2020}. The kite routine by \citet{Agol2020} provides both $a$ and $\kappa_0$ in a single call and needs to be evaluated at most once per exposure.
\end{enumerate}

The last approximation introduces minor errors during ingress and egress ($1 - k < b < 1 + k$), but its effect on inferred transit depths and orbital parameters is negligible for realistic $k$ variations. Numerical tests\footnote{\url{https://github.com/hpparvi/PyTransit/blob/master/notebooks/tsmodel/test_accuracy.ipynb}} show that for a hot Jupiter like WASP-39~b ($k$ varying from 0.144 to 0.154), the maximum transit depth error is 0.01\% (3 ppm), and for a sub-Neptune like TOI-836~c ($k$ from 0.033 to 0.036), the error is below 0.004\% (0.05 ppm). For grazing hot Jupiters, the error can rise to $\sim$0.5\%, but this remains below current observational precision and can be corrected if needed.

\subsection{Radius Ratio Parameterization} 
\label{sec:exoiris.radius_ratio}

In transmission spectroscopy, the key quantity of interest is the wavelength-dependent planet-star radius ratio, $k(\lambda)$, which encodes the shape of the transmission spectrum. While the \tsmodel enables the efficient calculation of thousands of spectrophotometric transit light curves, treating the radius ratio in each passband as an independent free parameter would result in an exceedingly large-dimensional model.

\exoiris addresses this by decoupling the data resolution from the spectral resolution of the retrieved transmission spectrum. As shown in Fig.~\ref{fig:model_example}, the radius ratio is modeled as an interpolating function over wavelength, defined by a user-specified set of knots. The values at the knot positions are treated as free parameters, and the radius ratio for each passband is interpolated using one of the implemented interpolation schemes.
Both the number and placement of the knots are fully customizable, and the locations of any individual knots can also be included as free model parameters. In typical use, knots can be distributed uniformly or logarithmically in wavelength, or clustered around known or suspected spectral features. Although the transit model is evaluated at the full spectroscopic data resolution, the radius-ratio values are constrained by the lower-dimensional interpolating function.

This parameterization offers several benefits. First, model complexity depends on the number of radius-ratio knots rather than on the data resolution. Second, the spectral resolution of the transmission spectrum can be tailored to the science case---for example, by placing more knots near expected absorption features and fewer in smooth continuum regions. Finally, the knot configuration can be modified dynamically during analysis, allowing for iterative refinement: an initial coarse (low-resolution) spectrum can be followed by a higher-resolution fit using the earlier result as a starting point.

\begin{figure*}
    \centering
    \includegraphics[width=1\linewidth]{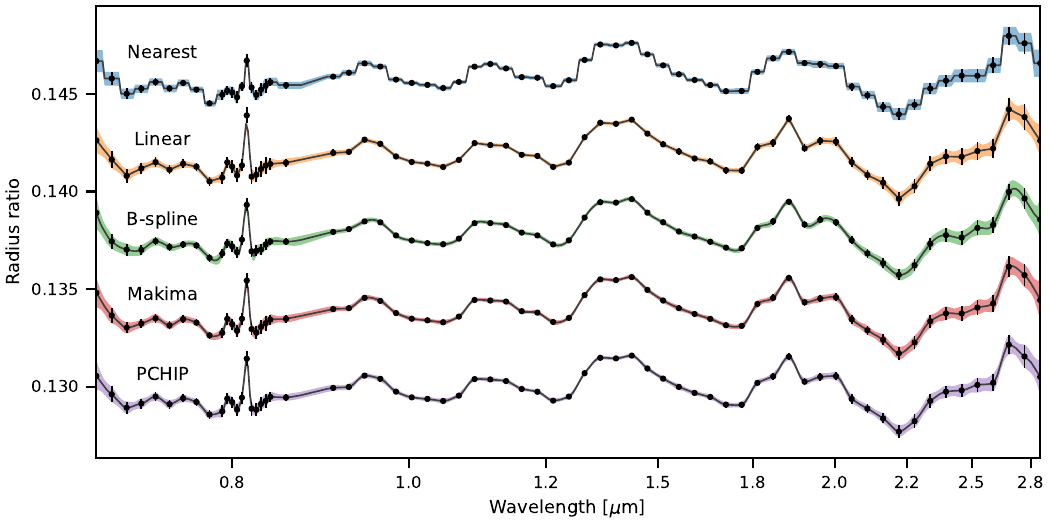}
    \caption{Radius-ratio interpolation schemes supported by \exoiris. The black dots show the posterior median radius-ratio knots with their 68\% central posterior intervals, as inferred from the \jwst WASP-39~b NIRISS observations presented in \citet{Feinstein2023} using five of the supported interpolation schemes (shifted in y for visualization). The black lines show the posterior median transmission spectra, and the colored shading shows their 68\% central posterior intervals.}
    \label{fig:interpolation_options}
\end{figure*}

Currently, \exoiris supports nearest-neighbor, linear, quadratic B-spline, cubic B-spline, PCHIP, and Makima interpolation (Fig.~\ref{fig:interpolation_options}), all implemented via \texttt{SciPy} \citep{Virtanen2020}. The nearest-neighbor scheme models the transmission spectrum as a piecewise constant step function, effectively treating the regions around knots as discrete bins, while the linear scheme connects the knots with straight line segments. The linear option is generally the most robust choice for low-signal-to-noise (SNR) datasets. The quadratic and cubic B-spline schemes produce smooth spline curves, but they can introduce spurious oscillations (overshoot) in low-SNR regions or near sharp spectral features. To mitigate this while retaining smoothness, \exoiris offers the PCHIP and Makima options. These are shape-preserving, non-linear interpolators that enforce monotonicity to prevent artificial extrema, though they are computationally more expensive than the other schemes.

\subsection{Limb Darkening}
\label{sec:exoiris.limb_darkening}

\subsubsection{Analytical Limb-Darkening Models}

\exoiris models the wavelength-dependent stellar limb-darkening coefficients using the same interpolation-based approach as for the planet-star radius ratio: a user-defined set of knots defines an interpolating function, and the coefficient values at these knots are treated as free parameters (see Fig.~\ref{fig:model_example}).

This parameterization is available for all two-parameter analytical limb-darkening laws supported by the \roadrunner model, including the quadratic and power-2 laws. It provides the same flexibility and regularization benefits as the interpolation-based radius-ratio model: the number and placement of knots can be adjusted to balance spectral resolution against model complexity, and the knot configuration can be updated at any point during the analysis.

As with the radius ratio, users can increase the knot density in regions where limb darkening is expected to change rapidly, or reduce the number of free parameters in smoother spectral regions. When applied to a joint fit of all spectrophotometric light curves, this approach reduces reliance on external priors and allows limb-darkening constraints to be derived empirically from the data. 
Although \exoiris supports the use of priors generated with the \ldtk package \citep{Parviainen2015b} (which uses PHOENIX stellar atmosphere models by \citealt{Husser2013} to generate theoretical limb-darkening profiles), these priors can be made broad enough to ensure that the inference is driven by the data, not the theory, when discrepancies arise.

\subsubsection{LDTk-Based Limb-Darkening Model}

In addition to the interpolation-based analytical models, \exoiris supports a numerical limb-darkening model based on \ldtk. This method uses PHOENIX stellar atmosphere models \citep{Husser2013} to generate intensity profiles directly from three stellar parameters: effective temperature ($\teff$), surface gravity ($\logg$), and metallicity (\metal). These parameters are treated as global free parameters, independent of the number of passbands, providing a low-dimensional and physically motivated alternative to spline interpolation.

While the \ldtk-based model is more tightly constrained by stellar theory, the interpolation-based approach offers greater flexibility when empirical, data-driven constraints are preferred. Both methods are fully supported in \exoiris and can be chosen based on the desired balance between theoretical and observational constraints.

\subsection{Additive Flux Offset}
\label{sec:exoiris.offset}

Errors in data reduction, such as incorrect bias estimation or instrumental systematics, can lead to apparent depth differences between detectors, as suggested for the NIRSpec observations of TOI-776~b \citep{Moran2023, Alderson2025}, where non-physical jumps were observed between the transmission spectra from the NRS1 and NRS2 detectors. The offset term, $O$, in Eq.~\eqref{eq:full_model} can be used to account for additive bias-level variations arising from instrumental or astrophysical sources. In general, these variations can be instrument- or epoch-dependent and directly affect the measured transit depths.

Each dataset is assigned to an \emph{offset group}, and a separate $\bias_i$ parameter is introduced for each group. Offset group~0 serves as the reference and has its offset fixed to zero, while the offsets for other groups are treated as free parameters. The offset $\bias$ modifies the overall transit depth uniformly across all wavelengths within a group, leaving the relative shape of the transmission spectrum unchanged.

This implementation allows the model to absorb detector- or epoch-specific flux shifts without distorting the spectroscopic features of interest. It also provides a useful diagnostic: non-zero offsets may indicate instrumental or astrophysical systematics at the group level. However, the method relies on sufficient wavelength overlap between offset groups, either directly with the reference group or indirectly through overlapping intermediate groups. Without such overlap, the offset becomes degenerate with the radius ratio, and additional constraints, such as informative priors on the offset parameters, are required to resolve the degeneracy. Alternatively, without wavelength overlap, the datasets can all be assigned to offset group 0 and the offset modeling can be carried out in the retrieval phase.

\begin{figure*}
    \centering
    \includegraphics[width=\linewidth]{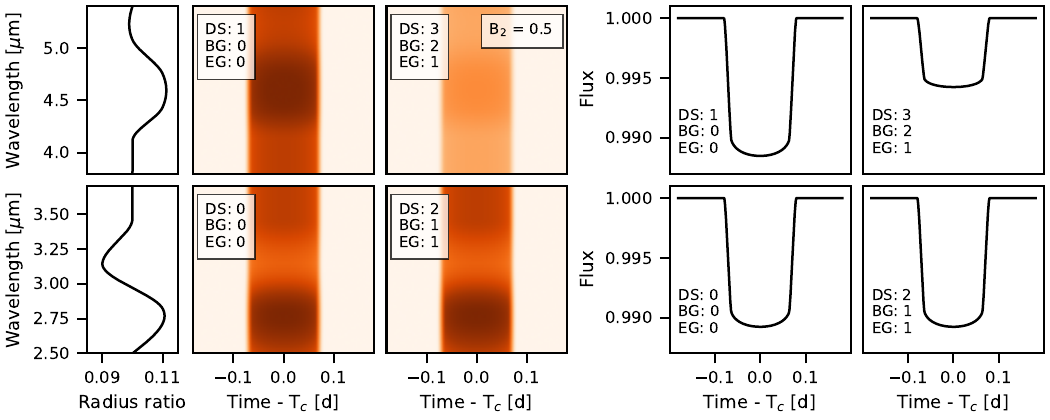}
    \caption{Example illustrating the use of offset groups to account for instrument- or epoch-dependent additive offsets arising from, for example, errors in bias-level estimation or instrumental systematics. The figure shows four spectrophotometric datasets from two transit epochs, each observed with two detectors. The first column shows the radius ratio as a function of wavelength, the next two columns show the spectrophotometric model for all datasets, and the final two columns display the average light curves. The dataset index is denoted by DS, the offset-group index by BG, and the epoch-group index by EG (see Sect.~\ref{sec:exoiris.ttvs}). Datasets 0 and 1 were observed at epoch 0 and do not feature an offset between the two detectors. However, datasets 2 and 3, observed at epoch 1, show an offset, with the transit of dataset 3 being, on average, half the transit depth of dataset 1.}
    \label{fig:bias}
\end{figure*}

Figure~\ref{fig:bias} illustrates this functionality using a scenario similar to that encountered by \citet{Alderson2025}, where two transits were observed with both the NRS1 and NRS2 detectors. In the second transit, a shift is visible between the NRS1 and NRS2 light curves, while no such shift appears in the first transit. Datasets~0 and~1 correspond to the first epoch and are assigned to the reference offset group (group~0). Datasets~2 and~3, from the second epoch, are assigned to offset groups~1 and~2, respectively, introducing two additional offset parameters. In the example shown, the model is evaluated with offset values of~0.0 and~0.5 for groups~1 and~2, resulting in identical flux models for datasets~0 and~2. However, the apparent transit depths for dataset~3 are reduced by~50\% relative to dataset~1, even though the physical radius ratio remains unchanged.

\subsection{Multiplicative Flux Baseline}
\label{sec:exoiris.baseline}

In addition to the additive offset term, the flux model includes a wavelength-dependent multiplicative baseline term, $B_\mathrm{d}(t, \lambda)$, which accounts for both the out-of-transit flux level and deterministic systematics. The baseline is modeled as a linear combination of dataset-specific covariates that, by default, include a constant intercept and a user-defined number of polynomial time terms centered on the dataset's mid-time. Furthermore, the user can incorporate any relevant covariates for each dataset as required. For example, common-mode systematics estimated from an initial white-light-curve fit (a common practice in ground-based transmission spectroscopy), or auxiliary variables observed simultaneously with the photometry.

The baseline coefficients are determined through analytic profiling: during each log-posterior evaluation, a linear least-squares fit is performed against the multiplicative residuals (defined as the observed flux divided by the current transit model) to identify the best-fit baseline for the proposed transit parameters. By analytically resolving the linear baseline coefficients, \exoiris ensures that baseline trends are self-consistently accounted for (and profiled out during posterior sampling) without increasing model dimensionality.

\subsection{Transit Timing Variations}
\label{sec:exoiris.ttvs}

Transit timing variations (deviations from strict periodicity in the transit center times, TTVs) are typically caused by gravitational interactions between planets in multi-planet systems. While TTVs are not an issue when analyzing transits individually, they must be accounted for in joint analyses of observations from multiple epochs. Failing to do so can distort the light curve shape and bias the inferred planet-star radius ratio, leading to inaccurate transmission spectra.

\exoiris accounts for TTVs by assigning each dataset to an epoch group. Each epoch group introduces a separate transit center time as a free model parameter. This allows datasets to be grouped by observation epoch, with TTVs automatically incorporated into the model. 

\begin{figure}
    \centering
    \includegraphics[width=\linewidth]{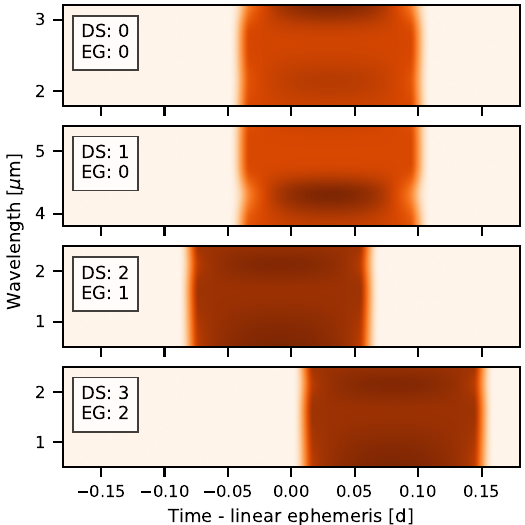}
    \caption{Example illustrating the use of epoch groups (EG) to account for transit timing variations in a joint transmission spectrum analysis with \exoiris. The example includes four spectrophotometric datasets: two from the same epoch observed with different instruments, and two from separate epochs observed with a third instrument.}
    \label{fig:ttvs}
\end{figure}

Figure~\ref{fig:ttvs} shows an example with four spectrophotometric datasets observed over three epochs. The first two datasets correspond to the same transit observed with different instruments, while the remaining two are from separate epochs and observed with a third instrument. In this setup, three transit center parameters are included in the model, one for each unique epoch index.

If TTVs are expected to be negligible, all datasets can be assigned to epoch group 0. However, assigning each transit epoch its own group adds minimal model complexity and can provide useful constraints. For high-precision datasets, the resulting transit center posteriors may reveal subtle TTV signals or help refine orbital parameters.

\subsection{Spot Crossings and the Transit Light Source Effect}
\label{sec:stellar_inhomogeneities}

Stellar spot crossings and flux contamination from unocculted stellar spots and faculae---also known as the transit light source effect \citep[TLSE;][]{Pont2008, Sing2011a, Pont2013,McCullough2014,Rackham2018,Rackham2019}---all affect the inferred transmission spectra. 

Individual stellar spot crossings are straightforward to identify, as these events produce characteristic bumps in the transit light curves. 
The TLSE arising from unocculted stellar inhomogeneities presents a more complex challenge. Both unocculted spots and faculae imprint wavelength-dependent signatures on the measured transit depths, with amplitudes and shapes determined by their temperatures and the stellar disk covering fractions.

\begin{figure}
    \centering
    \includegraphics[width=\linewidth]{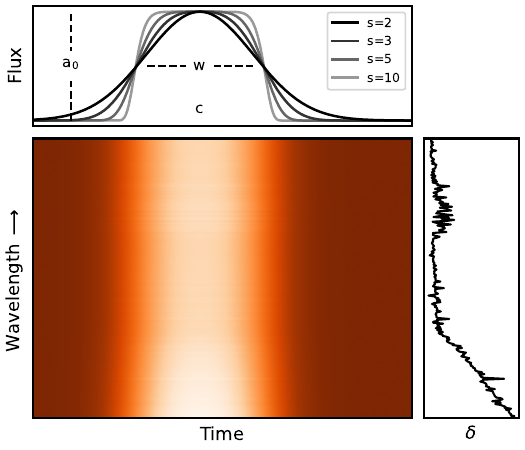}
    \caption{Model used for stellar spot crossing events. The temporal shape of each event is modeled as a generalized Gaussian, while the wavelength dependence of the spot amplitude is computed using BT-Settl model spectra by \citet{Allard2012a}.}
    \label{fig:spot_model}
\end{figure}

\exoiris allows an arbitrary number of stellar spots to be included in the flux model. Each spot is associated with an epoch group and is modeled as a generalized Gaussian (Fig.~\ref{fig:spot_model}),
\begin{multline}
    f(t, \lambda, T_\star, T_\mathrm{s}, c, a_0, w, s) = \\
    \delta(\lambda, T_\star, T_\mathrm{s}) a_0 \exp \left( -\frac{2\;|t-c|}{w\ln(4)^{1/s}} \right)^s,
\end{multline}
where $t$ is time, $\lambda$ is wavelength, $\delta$ is the wavelength-dependent spot contrast (normalized to unity at a reference wavelength $\lambda_0$), $T_\star$ and $T_\mathrm{s}$ are the photospheric and spot temperatures, respectively, $a_0$ is the spot amplitude at the reference wavelength, $w$ is the spot full width at half maximum, and $s$ is a shape parameter controlling the spot profile. The spot contrast, $\delta$, is computed from the stellar photosphere and spot temperatures using BT-Settl spectrum models \citep{Allard2012a}.

The effects from unocculted stellar heterogeneities can also be included into the joint model. Enabling TLSE modeling introduces the average temperatures of unocculted spots and faculae as free model parameters, and adds separate spot and faculae covering fractions as free parameters for each epoch group. The TLSE is then applied following the formalism by \citet{Rackham2018}, where the transit model is multiplied by
\begin{equation}
    \epsilon_{\lambda, \mathrm{s+f}} = \frac{1}{1 - 
    f_\mathrm{s}\left(1 - \frac{F_{\lambda, \mathrm{s}}}{F_{\lambda, \mathrm{p}}}\right) -
    f_\mathrm{f}\left(1 - \frac{F_{\lambda, \mathrm{f}}}{F_{\lambda, \mathrm{p}}}\right)},
\end{equation}
where $f_\mathrm{s}$ and $f_\mathrm{f}$ denote the fractional disk coverage of spots and faculae for each epoch, and $F_{\lambda, \mathrm{p}}$, $F_{\lambda, \mathrm{s}}$, and $F_{\lambda, \mathrm{f}}$ are the fluxes from the stellar photosphere, spots, and faculae, respectively.

\subsection{Noise Modeling}
\label{sec:exoiris.noise}

Robust noise modeling is important in transmission spectroscopy, particularly when combining observations from multiple epochs or instruments. \exoiris provides a flexible approach to noise modeling based on the concept of \emph{noise groups}, similar to the offset and epoch groups described earlier. Each spectrophotometric dataset is assigned to a specific noise group, allowing for tailored noise treatment across heterogeneous datasets.

\exoiris currently supports three noise modeling options: white noise, GPs with fixed hyperparameters, and GPs with free hyperparameters. GPs are implemented using the \texttt{celerite2} package \citep{Foreman-Mackey2023}.

The white noise model assumes independent, normally distributed uncertainties, scaled by a single multiplicative factor per noise group. These uncertainties are provided as an array with the same dimensions as the flux array, containing one value per flux measurement. They can be supplied by the data reduction pipeline or estimated during dataset initialization using the utility methods in \exoiris. The noise scaling factor is included as a free parameter to account for possible under- or overestimation of the input uncertainties. This model is appropriate when the noise is dominated by photon noise and there is minimal time-correlated or systematic variability. The log-likelihood is
\begin{equation}
\logl = -\frac{1}{2} \sum_j^{n_\mathrm{ds}} \sum_i^{n_{\mathrm{p}(j)}} \left(\log\left(2\pi s_{\mathrm{g}(j)}^2\sigma_{j,i}^2\right) + \frac{\left(f_{j,i} - m_{j,i}\right)^2}{s_{\mathrm{g}(j)}^2\sigma_{j,i}^2}\right),
\end{equation}
where $n_\mathrm{ds}$ is the number of datasets, $n_{\mathrm{p}(j)}$ the number of points in each dataset (the two dimensions are here unrolled to one for simplicity), $s_{\mathrm{g}(j)}$ is the noise multiplier for the noise group assigned to the $j$th dataset, $f_{j,i}$ is the observed flux, $m_{j,i}$ is the model flux, and $\sigma_{j,i}$ is the flux uncertainty.

The log-likelihood for a Gaussian Process (GP) noise model is 
\begin{equation}
\logl = -\frac{1}{2} \sum_j^{n_\mathrm{ds}} \left( n_{\mathrm{p}(j)} \log(2\pi)  + \log\left|\covmat_j\right| + \vec{r}_j^{\mathsf{T}} \covmat_j^{-1} \vec{r}_j \right),
\end{equation}
where $\vec{r}_j = \vec{f}_j - \vec{m}_j$ is the residual vector containing the observed fluxes $\vec{f}_j$ and model fluxes $\vec{m}_j$. The covariance matrix, \covmat, is constructed as
\begin{equation}
    \covmat = \mathbf{K(t,t)} + \boldsymbol{\sigma}^2\mathbf{I},
\end{equation}
where $\mathbf{K}$ is the covariance kernel with either fixed or fitted hyperparameters and $\vec{t}$ are the mid-exposure times. The GP noise model considers only time-dependent systematics.

The GP model with fixed hyperparameters uses a user-defined kernel to describe correlated noise or long-term trends. The hyperparameters can be defined separately for each noise group, but they cannot vary along the wavelength or time axes. Any kernel supported by \texttt{celerite2} can be used, and \exoiris includes methods for optimizing hyperparameters for the Mat\'ern-3/2 kernel. This approach is suitable in most cases where correlated noise or longer-time-scale trends are present, and it offers a good balance between flexibility and computational efficiency.

The GP model with free hyperparameters allows the hyperparameters to be estimated jointly with the rest of the model parameters. Currently, this is supported only for the basic Mat\'ern-3/2 kernel and comes with a higher computational cost. Support for additional kernels may be added in future versions.

The GP is evaluated independently for each wavelength channel using \texttt{celerite2}. Internally, this is implemented by reshaping each two-dimensional residual array of size $(n_{\mathrm{wl}}, n_{\mathrm{time}})$ into a one-dimensional array of length $n_{\mathrm{wl}} \times n_{\mathrm{time}}$. The corresponding time array is repeated $n_{\mathrm{wl}}$ times, and a constant time offset is added between successive repetitions such that different passbands are separated by a sufficiently large temporal gap. This ensures that the covariance between points belonging to different passbands is effectively zero. The GP log-likelihood for all passbands is then evaluated in a single call using the combined time array and the flattened residuals.

Although a single \exoiris analysis cannot combine different noise models, individual noise groups can use different GP kernels or kernel parameters. Additionally, users can change the noise model during an analysis or continue from a saved solution. This allows, for example, a quick initial fit using a white noise model to be followed by a more detailed analysis with a GP noise model.

\subsection{Masking of Bad Data}
\label{sec:exoiris.masking}

Each \exoiris dataset includes a mask array that flags spectrophotometric data points to be excluded from the log-likelihood evaluation. Non-finite values, such as NaNs and infinities, in the flux or uncertainty arrays are automatically masked. In addition, \exoiris provides tools to detect and mask outliers, allowing users to remove problematic data points before or during the analysis.

\subsection{Optimization and Posterior Inference}
\label{sec:exoiris.inference}

The transmission spectrum inference in \exoiris is divided into a global optimization phase and a subsequent posterior sampling phase. The first phase uses a Differential Evolution (DE) optimizer \citep{Storn1997, Price2005} to identify the region of the global posterior mode. The optimization is initialized with a population of $n$ parameter vectors, and, during the optimization, the DE algorithm concentrates the parameter vector population toward the global posterior mode.

Transition to the sampling phase occurs once the population has concentrated into a sufficiently tight clump. Convergence is determined by the width of the log-posterior distribution within the population, and the convergence limit is user-adjustable. Posterior inference is then performed using the affine invariant Markov Chain Monte Carlo (MCMC) sampler implemented in the \texttt{emcee} package \citep{Foreman-Mackey2012}. By initializing the sampler with the clumped population from the optimization phase, the Markov chains start their exploration near the global posterior mode. 

Both optimization and sampling support parallel execution via Python's multiprocessing module or the \texttt{mpi4py} package.
Additionally, both optimization and MCMC sampling can be performed semi-interactively. The optimizer can be executed for a specific number of iterations, after which the current best-fit model and population state can be examined using various diagnostics. The optimization can then be resumed for additional iterations, continuing directly from its previous state without loss of progress. Similarly, the MCMC sampler can be run for a set number of iterations to permit the inspection of intermediate results, such as chain convergence and autocorrelation times, before continuing sampling from the last recorded state.

The state of an \exoiris analysis can be saved at any time. Every saved model is fully self-contained because the save file stores all information required to reproduce the analysis, including datasets, priors, and the current state of the parameter population. This architecture enables iterative and branching workflows in which a saved analysis serves as the basis for subsequent analyses. Users can introduce additional model complexity, such as adding radius-ratio knots, freeing knot locations, incorporating the transit-light-source effect, or updating datasets, while leveraging the progress made in earlier stages of the analysis.

\subsection{Atmosphere Retrieval with Oversampled Spectra}
\label{sec:exoiris.reduced_likelihood}

After sampling, the \texttt{ExoIris.transmission\_spectrum} utility method can be used to generate a posterior transmission spectrum. This method computes the mean transmission spectrum, \meansp, and its covariance matrix, \covmat, on an arbitrary wavelength grid with $n_\lambda$ elements. Additionally, the \texttt{ExoIris.create\_loglikelihood\_function} method can be used to create a callable \texttt{LogLikelihood} object ready for use in atmospheric retrieval frameworks.

A specialized approach to calculating the log-likelihood is required because the transmission spectrum is oversampled. That is, the transmission spectrum is derived by interpolating the posterior distribution of $n_\mathrm{k}$ radius-ratio knots, where, typically, $n_\mathrm{k} \ll n_\lambda$, and the transmission spectrum elements in \meansp are strongly correlated across wavelengths. Consequently, the points cannot be treated as independent observations, and the full covariance structure must be accounted for when evaluating the likelihood of theoretical atmospheric models.

Ideally, a theoretical transmission spectrum $\vec{m}$ would be compared to the posterior using the standard multivariate normal log-likelihood,
\begin{equation}
    \logl = -\frac{1}{2} \left(n_\lambda \log(2\pi) + \log(|\covmat|) + \vec{r}^\mathsf{T} \covmat^{-1} \vec{r}\right),
\end{equation}
where $\vec{r} = \meansp - \vec{m}$ is the residual vector. However, the interpolation process renders the posterior transmission spectrum rank-deficient. For linear interpolation methods (nearest-neighbor, linear, and B-spline), the rank of \covmat is strictly limited to $n_\mathrm{k}$ instead of $n_\lambda$ because the spectrum at any wavelength is a linear combination of the $n_\mathrm{k}$ knot values. For non-linear interpolators (\texttt{PCHIP}, \texttt{Makima}), while the algebraic rank may exceed $n_\mathrm{k}$, the effective rank still remains constrained to $\approx n_\mathrm{k}$. This rank deficiency implies that \covmat is ill-conditioned and that its direct inversion is numerically unstable (or plain impossible), rendering the standard likelihood evaluation infeasible.

To construct a robust likelihood, \exoiris adopts the subspace compression formalism, often referred to as Karhunen-Lo\`eve (KL) compression, widely used in cosmology to handle rank-deficient datasets \citep[e.g.,][]{Tegmark1997}. The code performs an eigendecomposition of the empirical covariance,
\begin{equation}
    \covmat = \mathbf{U} \, \boldsymbol{\Lambda} \, \mathbf{U}^{\mathsf{T}},
\end{equation}
and identifies the eigenvalues that are significantly larger than numerical noise. Only these eigenvalues and their corresponding eigenvectors represent directions in which the posterior distribution has meaningful support. Eigenvalues smaller than a given tolerance are discarded. The retained eigenvectors, $\mathbf{U}_r$, form an orthonormal basis for the reduced subspace, and the retained eigenvalues (variances along the eigenvectors) form the diagonal matrix $\boldsymbol{\Lambda}_r$.

The likelihood is then evaluated by projecting the residuals onto the subspace defined by the eigenvectors. The resulting subspace log-likelihood is
\begin{equation}
    \logl = -\frac{1}{2} \left( K \log(2\pi) + \sum_{i=1}^{K} \log \lambda_i + \sum_{i=1}^{K} \frac{(\vec{u}_i^\mathsf{T} \vec{r})^2}{\lambda_i} \right),
\end{equation}
where $\vec{u}_i$ is the $i$th eigenvector, $\lambda_i$ is the $i$th eigenvalue, and the summation runs only over the $K$ valid eigenvectors.

\section{Example: WASP-39~b Early Release Science Joint Analysis}
\label{sec:example}

The \exoiris example GitHub repository\footnote{\url{https://github.com/hpparvi/exoiris_examples}} contains a set of examples ranging from simple analyses with a single dataset to more complex cases combining multiple datasets into a joint analysis. The WASP-39~b Early Release Science (ERS) joint analysis provides a comprehensive demonstration of a joint transmission spectrum analysis using spectrophotometric light curves from WASP-39~b transits observed with NIRSpec, NIRSpec PRISM, NIRCam, and NIRISS. The data used in this example were produced by \citet{Carter2024}, who gathered the original observations presented by \citet{Ahrer2023}, \citet{Feinstein2023}, \citet{Alderson2023}, and \citet{Rustamkulov2023}. The example illustrates how to configure bias, ephemeris, and noise groups for joint modeling and includes individual transmission spectroscopy analyses for each instrument.

\subsection{Analysis Setup}

\begin{figure*}
    \centering
    \includegraphics[width=1\linewidth]{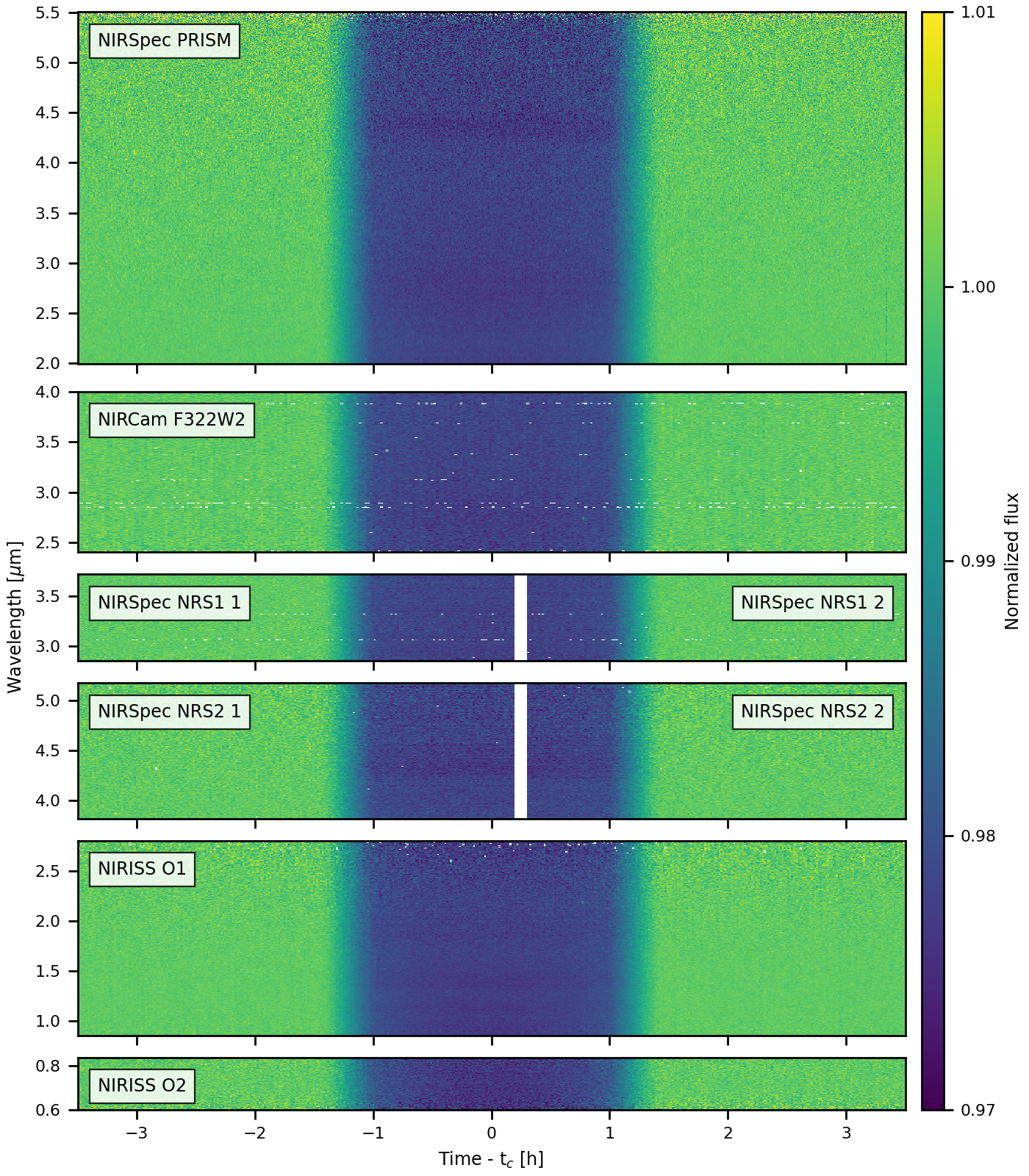}
    \caption{Spectrophotometric light curves used in the WASP-39~b joint analysis. White pixels indicate outliers that were masked from the analysis.}
    \label{fig:w39_data}
\end{figure*}

\begin{figure*}
    \centering
    \includegraphics[width=1\linewidth]{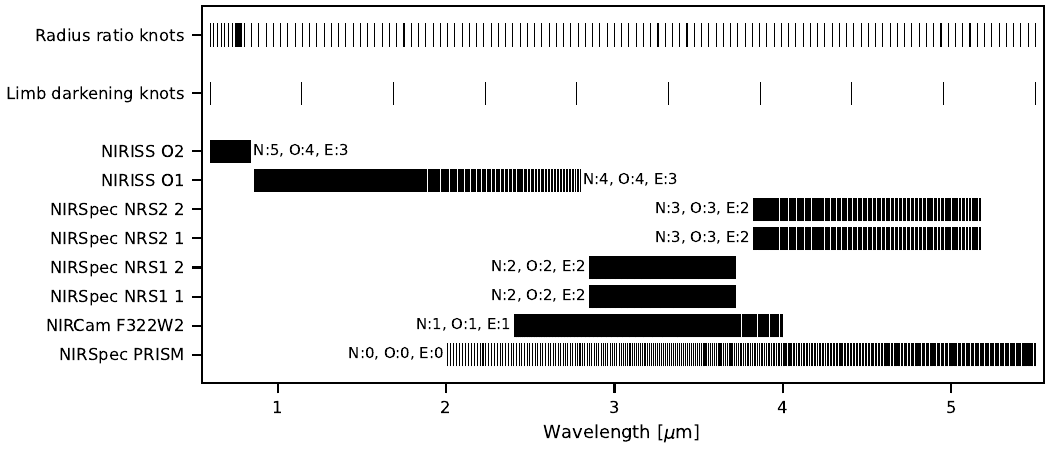}
    \caption{\exoiris setup for the WASP-39~b ERS joint analysis. The top row shows the radius-ratio knot locations, the second row shows the limb-darkening knot locations, and the remaining rows show the central wavelengths of the spectroscopic elements for each dataset. The noise (N), offset (O), and epoch (E) group assignments are indicated beside each dataset’s wavelength coverage. Details on cropping and binning are provided in the text.}
    \label{fig:w39_setup}
\end{figure*}

The joint analysis includes eight spectrophotometric datasets shown in Fig.~\ref{fig:w39_data}. The NIRSpec PRISM data are cropped to $2.0 \leq \lambda \leq 5.5$~\micron to remove saturated wavelengths \citep{Carter2024, Rustamkulov2023}, and then binned in time to 30~s (but not binned in wavelength). The NIRCam F322W2 data are cropped to $\lambda < 4$~\micron and $t > 2459783.32$, and binned in wavelength to a resolution of $R \approx 600$. The NIRSpec G395H data are split into two time segments ($2459791.45 < t < 2459791.62$ and $t > 2459791.625$), cropped to $\lambda > 2.85$~\micron, and binned in wavelength to $R \approx 600$, resulting in four datasets. The NIRISS SOSS order 1 data are cropped to $0.86 \leq \lambda \leq 2.8$~\micron and binned to $R \approx 300$, while order 2 data are cropped to $0.60 \leq \lambda \leq 0.84$~\micron and binned to $R \approx 800$. Instead of using uncertainties from the reduction pipelines, I estimate the white-noise scatter (standard deviation) for each light curve directly from the data.

Each of the four transit observations is assigned to a separate epoch group, adding four transit center parameters to the model. Although WASP-39~b shows no signs of TTVs, the \citet{Carter2024} NIRISS reduction produces a significant offset in transit center time relative to a linear ephemeris, in contrast to the other datasets. While this transit time offset is a data-reduction artifact, it demonstrates the functionality of epoch groups in \exoiris. Each detector (and NIRISS SOSS order) is assigned a separate bias offset group, resulting in five free offset parameters. The NIRSpec NRS1 dataset is used as the reference (offset group 0). Each dataset is also assigned to its own noise group, resulting in six white-noise scaling parameters. The final model setup is illustrated in Fig.~\ref{fig:w39_setup}.

The transmission spectrum estimation is carried out progressively, starting with a model that uses nearest-neighbor radius-ratio interpolation and 25 radius-ratio knots, $n_\mathrm{k}$. After fitting and sampling the initial model, the model resolution is increased to $n_\mathrm{k}=60$, $n_\mathrm{k}=130$, and $n_\mathrm{k}=230$, every time saving the models and the posterior sampling results separately. The region around the 0.768~\micron K line is sampled with a higher radius-ratio knot resolution after the initial run. The analyses are repeated for the linear, quadratic B-spline, cubic B-spline, PCHIP, and Makima radius-ratio interpolation schemes. Limb darkening is modeled using 10 linearly spaced knots from 0.6 to 5.5~\micron. In the initial low-resolution analysis, the limb-darkening coefficients are constrained with loose, informative priors derived using \ldtk, whereas in subsequent analyses the priors are changed to uninformative uniform priors.

\subsection{Results}
\label{sec.example.results}

\begin{figure*}
    \centering
    \includegraphics[width=1\linewidth]{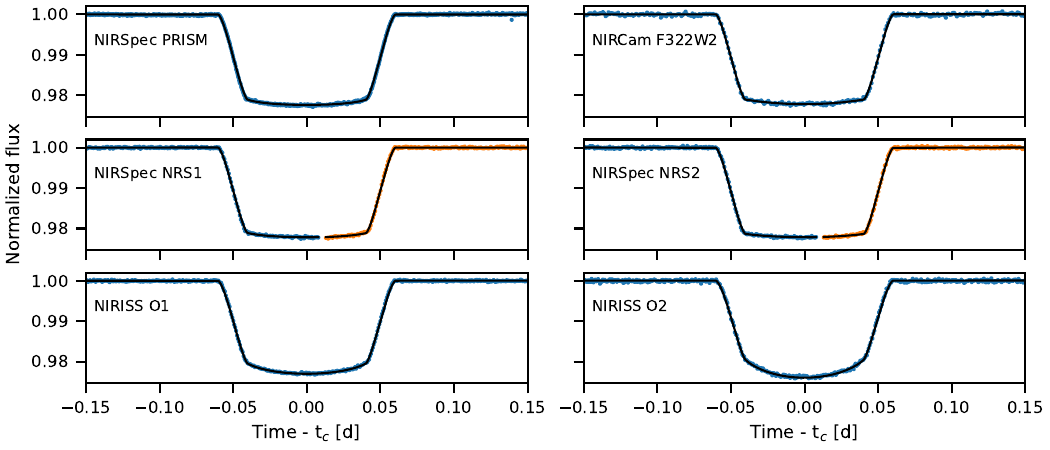}
    \caption{WASP-39~b white-light curves jointly fitted with \exoiris. Blue and yellow points show the observed photometry, and the black line shows the median posterior model. The NIRSpec NRS1 and NRS2 light curves are split in time due to a jump near mid-transit.}
    \label{fig:w39_white_lcs}
\end{figure*}

\begin{figure*}
    \centering
    \includegraphics[width=1\linewidth]{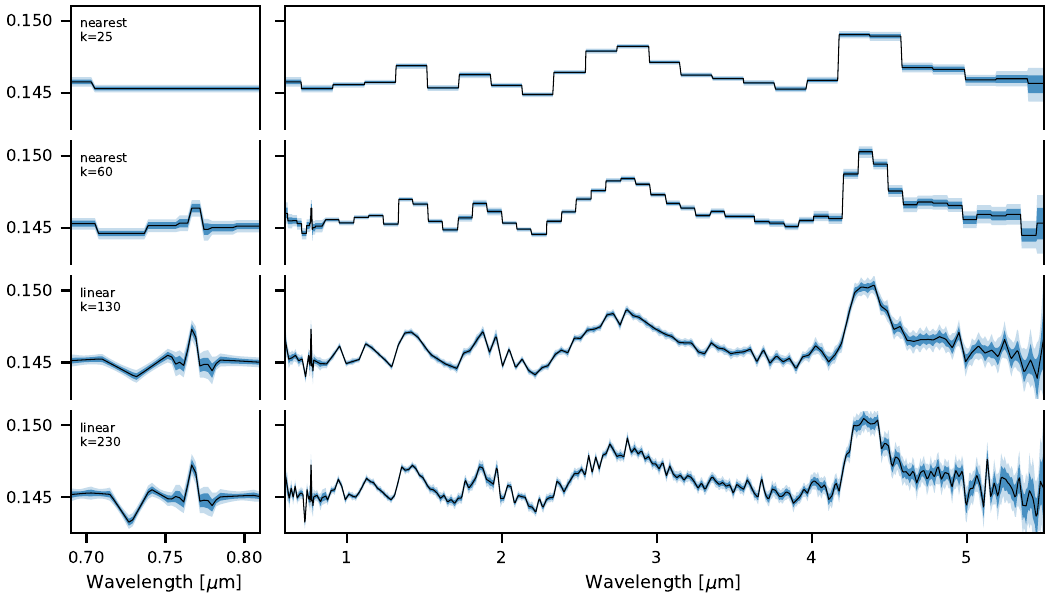}
    \caption{Posterior WASP-39~b transmission spectra estimated from the joint analysis of NIRSpec, NIRSpec PRISM, NIRCam, and NIRISS data using \exoiris. The black line shows the posterior median model, and the blue shading shows the 68\% and 95\% central posterior intervals. The transmission spectrum was estimated progressively, starting with nearest-neighbor interpolation with a low radius-ratio knot resolution (top), and ending with a linear interpolation with a high knot resolution (bottom).}
    \label{fig:w39_joint_spectrum_progression}
\end{figure*}

\begin{figure*}
    \centering
    \includegraphics[width=1\linewidth]{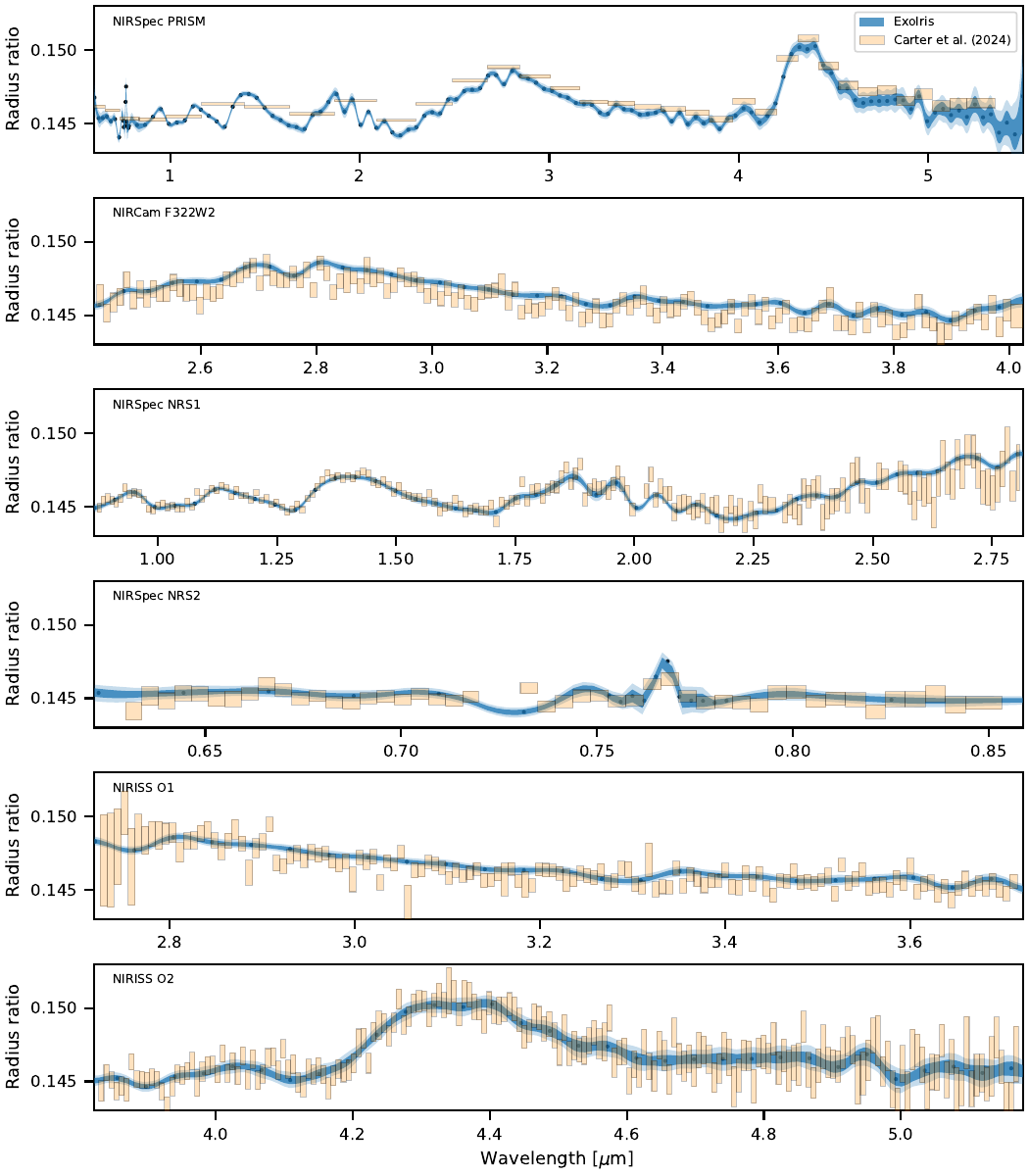}
    \caption{As in Fig.~\ref{fig:w39_joint_spectrum_progression}, but comparing the \citet{Carter2024} results with the \exoiris analysis using quadratic B-spline interpolation with 130 radius-ratio knots. The black points show the median posterior radius-ratio knot values, and the orange shading shows the $1\sigma$ intervals from the \citet{Carter2024} analysis with fitted limb darkening and bin scale 5.}
    \label{fig:w39_joint_spectrum}
\end{figure*}

\begin{figure}
    \centering
    \includegraphics[width=1\linewidth]{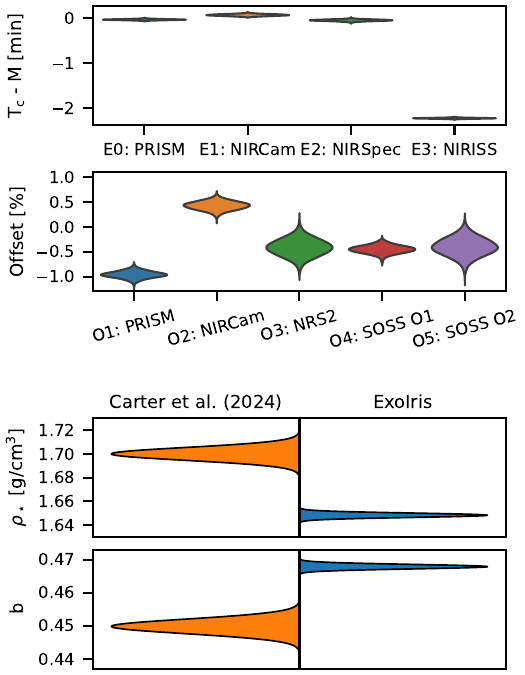}
    \caption{Posterior distributions for a subset of \exoiris model parameters from the WASP-39~b joint analysis. The first panel from the top shows the transit center time offsets for the four epoch groups, the second panel shows the additive offset posteriors, and the two bottom panels compare the stellar density and impact parameter posteriors to the estimates presented in \citet{Carter2024}.}
    \label{fig:w39_posteriors}
\end{figure}

Figure~\ref{fig:w39_white_lcs} shows a joint fit to the white-light curves. This is an optional initial step that can be used to initialize the transmission spectroscopy analysis, and is also useful for visual identification of spot crossings. Figure~\ref{fig:w39_joint_spectrum_progression} shows the WASP-39~b transmission spectrum estimates from four progressively complex joint analyses, and Fig.~\ref{fig:w39_joint_spectrum} compares the \exoiris-estimated transmission spectrum with quadratic B-spline radius-ratio interpolation and $n_\mathrm{k}=130$ against the results by \citet{Carter2024}. Finally, Fig.~\ref{fig:w39_posteriors} shows the posterior distributions for a set of \exoiris model parameters.

The \exoiris transmission spectrum agrees well with the \citet{Carter2024} spectra. The differences in the NIRSpec PRISM and NIRCam mean levels (Fig.~\ref{fig:w39_joint_spectrum}) are due to the instrument-dependent offsets discussed in \citet{Carter2024}. These offsets are captured by the \exoiris analysis, as shown in Fig.~\ref{fig:w39_posteriors}. Additionally, as mentioned in \citet{Carter2024}, \jwst observations offer unprecedented precision for transit-based stellar density estimation (assuming the eccentricity and argument of periastron are also known to high precision). Joint modeling of multiple spectrophotometric datasets naturally increases precision beyond that of individual analyses. Assuming a circular orbit, the \exoiris analysis yields a stellar density of $1.648 \pm 0.002$ and an impact parameter of $0.4679 \pm 0.0006$.

\section{Discussion}
\label{sec:discussion}

The \exoiris package provides a flexible and efficient framework for analyzing spectrophotometric time-series data in the context of exoplanet transmission spectroscopy. Its core design principles (direct modeling of two-dimensional spectrophotometric datasets and joint analysis of multi-instrument, multi-epoch observations) enable robust and self-consistent inference of both wavelength-dependent and wavelength-independent parameters.

A key strength of \exoiris lies in its explicit parameterization of wavelength-dependent quantities, such as the planet-star radius ratio and limb darkening, using interpolating models. This approach decouples the spectral resolution of the transmission spectrum from the data resolution, allowing users to balance model complexity against scientific goals. In addition, the inclusion of group-based structures for instrumental offsets, noise properties, and transit timing variations enables accurate joint modeling of heterogeneous datasets.

The example analysis of WASP-39~b highlights the effectiveness of these methods. \exoiris accommodates diverse instrument modes, detector configurations, and data characteristics within a single modeling framework. The results reproduce the transmission spectrum derived by \citet{Carter2024} and provide improved flexibility for sensitivity testing, including the impact of systematic offsets and timing discrepancies. 

Additionally, \exoiris is computationally efficient. The 24 WASP-39~b joint analyses ($n_\mathrm{k}$ of 25, 60, 130, and 230, repeated for nearest-neighbor, linear, quadratic B-spline, cubic B-spline, PCHIP, and Makima radius-ratio interpolation) were all carried out during a single day using a MacBook Pro with an M3 Max CPU. High computational efficiency combined with the ability to perform iterative and branching workflows allows for an exploratory approach where the users can test hypotheses and refine models dynamically.

Finally, \exoiris remains under active development, driven by community feedback. Future updates will focus on further increasing the adaptability of the interpolation-based transmission spectrum model and improving the efficiency of the optimization and posterior sampling routines.

\begin{acknowledgments}
I acknowledge support by the Spanish Ministry of Science and Innovation with the Ram\'on y Cajal fellowship number RYC2021-031798-I.
Funding from the University of La Laguna and the Spanish Ministry of Universities is acknowledged.
\end{acknowledgments}

\software{PyTransit \citep{Parviainen2015, Parviainen2020a, Parviainen2020b},  
          LDTk \citep{Parviainen2015b}, 
          Astropy \citep{astropy:2013, astropy:2018, astropy:2022},
          emcee \citep{Foreman-Mackey2012},
          celerite2 \citep{Foreman-Mackey2023},
          NumPy \citep{Harris2020},
          SciPy \citep{Virtanen2020},
          Numba \citep{Lam2015}.
          }

\bibliography{tsmodel}{}
\bibliographystyle{aasjournal}
\end{document}